\documentclass[showpacs,prb,aps]{revtex4}
\usepackage{amssymb}

\usepackage{graphicx}

\begin{document}

\title{Decoherence of electron spin qubits in Si-based quantum computers}
\author{Charles Tahan, Mark Friesen, and Robert Joynt}
\affiliation{Department of Physics, University of Wisconsin-Madison\\
1150 University Avenue, Madison, WI 53706}
\date{\today }

\begin{abstract}
Direct phonon spin-lattice relaxation of an electron qubit bound by a donor
impurity or quantum dot in SiGe heterostructures is investigated. The aim is
to evaluate the importance of decoherence from this mechanism in several
important solid-state quantum computer designs operating at low
temperatures. We calculate the relaxation rate $1/T_1$ as a function of
[100] uniaxial strain, temperature, magnetic field, and silicon/germanium
content for Si:P bound electrons. The quantum dot potential is much
smoother, leading to smaller splittings of the valley degeneracies. We have
estimated these splittings in order to obtain upper bounds for the
relaxation rate. In general, we find that the relaxation rate is strongly
decreased by uniaxial compressive strain in a SiGe-Si-SiGe quantum well,
making this strain an important positive design feature. Ge in high
concentrations (particularly over 85\%) increases the rate, making Si-rich
materials preferable. We conclude that SiGe bound electron qubits must meet
certain conditions to minimize decoherence but that spin-phonon relaxation
does not rule out the solid-state implementation of error-tolerant quantum
computing.
\end{abstract}

\maketitle

\section{Introduction}

\label{sec:introduction}

The prospect of quantum computing (QC) has caused great excitement in
condensed matter physics. If a set of qubits can be maintained in a
coherent, controllable many-body state, certain very difficult computational
problems become tractable. In particular, successful QC would mean a
revolution in the areas of cryptography \cite{shor} and data-base searching %
\cite{grover}. \ In addition, it would mean a great advance in general
technical capabilities, since the control of individual quantum systems and
their interactions would represent a new era in nanotechnology.

However, from a practical point of view, a dilemma presents itself
immediately. On the one hand, one wishes to control quantum degrees of
freedom using external influences, since that is how a quantum algorithm is
implemented, and to measure them, since that is the output step. On the
other hand, the system must be isolated from the environment, since random
perturbations will destroy the quantum coherence that is the whole advantage
of QC. This is the isolation-control dilemma, and it leads to a very rough
figure-of-merit $F$ for any quantum computer. If we define the decoherence
time $\tau $ as the time it takes to lose quantum coherence, and the clock
speed $s$ (roughly the inverse of the time to run a logic gate), then the
figure of merit is $F=s\tau $ and a practical machine should satisfy $%
F>10^{4}-10^{5}$ at least. If the clock speed is limited only by standard
electronics, then we may be able to achieve $s\approx 10^{9}~Hz$. This would
imply that $\tau =1~ms$ is a lower limit for the decoherence time. \ 

The dilemma has not yet been solved, though a number of solutions have been
proposed. \ A particularly attractive solution is to use spin degrees of
freedom as qubits. \ Nuclear spins interact relatively weakly with their
environment because the coupling, proportional to the magnetic moment, is
small. \ Yet there has grown up a sophisticated technology (NMR) for the
manipulation of the spins, and some rudimentary computations have been
performed \cite{chuang}. \ Readout is the main difficulty with this
approach, since the field created by a single moment is tiny, and pure
states cannot be achieved in the macroscopic samples used.

Electron spins also interact weakly with the environment in some
circumstances: relaxation times in excess of $10^{3}~s$ have been measured
for donor bound states of phosphorus-doped silicon (Si:P) \cite{feher}. The
corresponding manipulation technology (ESR) has also reached a high level of
sophistication, but the magnetic moment of the electron exceeds that of the
nucleus by 3 orders of magnitude, which presents problems of isolation. \
Readout should be easier than for nuclei, since detection of single electron
charges is certainly possible \cite{kastner}, and spin-dependent detection
is not far out of the reach of current technology.

Solid-state implementations of QC are particularly attractive because of the
possibility of using existing computer technology to scale small numbers of
qubits up to the $10^{5}$ or so that would be needed for nontrivial
computations. \ The first paper to propose using the electron spin in a
quantum dot subjected to a strong DC magnetic field was that of Loss and
DiVincenzo \cite{loss}. \ Kane \cite{kane} proposed employing the nuclear
spin in the Si:P system as the qubit. \ A specific structure consisting of
silicon-germanium (SiGe) layers was proposed by Vrijen \textit{et al.} \cite%
{vrijen}. This structure incorporates the idea that the $g$-factor of an
electron can be changed by moving it in a Ge concentration gradient,
allowing individual electron to be addressed by the external AC field. \ A
different SiGe structure has been proposed by Friesen \textit{et al.} \cite%
{eriksson}. This structure is designed so that the electron number on the
dots, and the coupling between the dots, can be carefully controlled. \ 

Solid-state implementations must also face the isolation-control dilemma. \
Decoherence times must exceed the $1~ms$ number in the actual physical
structures that are needed for the operation of quantum algorithms. \ In
this paper, we examine whether this can be the case for some of the existing
proposals based on electron-spin qubits. \ In the process, we hope to learn
something about modifications to these structures that can increase $\tau $.
\ We shall focus on low temperature operation, since, as we shall see, this
will probably be necessary in order to obtain sufficiently large $\tau $. \ 

We can build on a large body of work, both theoretical and experimental,
from the 1950's and 1960's on ESR in doped semiconductors. \ In a series of
papers, Feher and collaborators \cite{feher} investigated the relaxation
time for the spin of electrons bound on donor sites in lightly doped Si. \
At sufficiently low temperatures, the relaxation time $T_{1}$ is dominated
by single-phonon emission and absorption. In the presence of spin-orbit
coupling (SOC), this can relax the spin, causing decoherence. \ The theory
was worked out by Hasegawa \cite{hasegawa} and Roth \cite{roth}.\ 

It must of course be recognized that this spin-lattice relaxation time is
not necessarily to be identified with the decoherence time. \ The
decoherence time is the shortest time for any process to permanently erase
the phase information in the wavefunction. \ This may mean the phase for a
single spin, but it also means that the relative phases of the wavefunctions
of different spins must also be preserved, so that processes that cause
mutual decoherence must also be taken into account. \ The actual decoherence
time is the minimum of all of these times. \ A spin relaxation time in
excess of $1~ms$ is a necessary, not a sufficient, condition for the
viability of a solid-state electron-spin QC proposal. \ 

A QC must have precise input as well as an accurate algorithm. \ Preparation
of the spin state is often proposed to be done by thermalization of the spin
system at a low temperature. \ The time to do this actually sets an \textit{%
upper} limit on the relaxation time of whatever processes thermalize the
spins to the lattice. \ A limit of perhaps $1-10$ $s$ is a reasonable
requirement.

This paper focuses on $T_{1}$, the time for relaxation of the longitudinal
component of the magnetization, by spin-phonon interactions. \ These
processes cause real spin-flip transitions. \ They occur at random times and
thus indubitably cause decoherence. \ In addition to these processes
characterized fully by $T_{1}$, there are processes which introduce random
phase changes in the spin wavefunctions. \ To characterize all such
processes by a single ''dephasing time'' $T_{2}$ will usually not be
sufficient for understanding the operation of a multi-qubit system \cite{hu1}%
. Difficulties of definition arise, and care must be taken to specify which
phase is involved and to what extent it is randomized. \ For a single spin
system there is no ambiguity. \ The $2\times 2$ density matrix $\rho _{ij%
\text{ }}$for the qubit with cylindrical symmetry involves only 2
independent parameters $\rho _{11}-\rho _{22}$, and $\rho _{12}.$ \ The time
dependence of $\rho _{11}-\rho _{22}$ after a system preparation is
exponential with decay constant $T_{1}$ and represents the return of the
longitudinal component of the magnetization to its equilibrium value. \ The
decay is due to inelastic transitions of the type calculated in this paper.
\ $\rho _{12}$ on the other hand, is nonzero only if the preparation of the
spin state has a transverse component: $S_{x}(t=0)\neq 0.$ \ The decay of
this quantity represents the irreversible conversion of this state to an
incoherent mixure of ''up'' and ''down'' states. \ Again, this is genuine
decoherence of the spin state, since the phase information cannot be
recovered. \ The time dependence of $\rho _{12}$ when the spin is in a
strong field has been calculated by Mozyrsky \textit{et al.} \cite{mozyrsky}
using a Markovian approximate master equation. \ They find that the time
scale of the decay due to spin-phonon coupling is very short, of the order
of the time for a phonon to cross the electron's wavefunction, which is
about $10^{-10}s$. \ But the decay is incomplete, with $\rho _{12}$
retaining all but $10^{-8}$ of its original value. Their calculation was for
Si:P, \ but a very similar result should hold for the dot case. \
Decoherence at this level is certainly acceptable for quantum computing. \
These authors also pointed out that the decay of the remainder of $\rho _{12}
$ is due to the spin-flip processes computed in this paper. \ If $T_{2}$ is
defined as the dominant decay time of the off-diagonal density matrix
element, then $T_{1}=T_{2}$ for spin-phonon processes.

A quite different source of decoherence is the hyperfine coupling to nuclear
spins. \ The nuclear spins produce an effective random magnetic field on the
electrons. \ Very recent calculations using semiclassical averaging
techniques \cite{efros} have obtained a very short relaxation time $%
T_{\Delta }\approx 1~ns$ for GaAs-based dot systems in a strong field. This
represents the decay of the transverse magnetization of an ensemble of dots.
\ This is a dephasing time, but \textit{not} a decoherence time. \ The
electrons spins precess in what is effectively the frozen field of the
nuclei. \ This field is spatially random, and the differential precession of
the electron spins leads to the magnetization decay. \ However, this is not
an irreversible loss of the phase information of the collective
wavefunction. \ Spin echo exeriments are a very beautiful demonstration of
precisely this point. This ''inhomogeneous broadening'' presents challenges
for the calibration and operation of quantum computers, but does not destroy
coherence.

Finally, in any implementation based on electron spin qubits, there will
certainly exist small interactions between the spins themselves. The
dipole-dipole interaction, for one, cannot be avoided, and there may be
indirect spin-spin interactions mediated by the gates. A recent paper
suggests that these interactions set the fundamental time scale $T_{M}$ for
Si quantum dot implementations of QC \cite{desousa}. These interactions do
produce experimental broadening of ESR lines in experiments on bulk systems,
and this might be taken as decoherence. In our view, however, these
interactions do not destroy the coherence of a state. The system is the set
of all the qubits. During the course of a quantum algorithm they are
collectively in a pure state (in principle). Any decoherence that destroys
the purity of the state comes from averaging over the unknown states of the
environment. The broadening that comes from dipole-dipole interactions
comes, in NMR and ESR calculations, from averaging over the states of the
system itself, which is not an appropriate method for calculating
decoherence. The effect of qubit-qubit interactions that cannot be turned
off is to complicate the quantum algorithm. A quantum algorithm is a unitary
transformation that must always include the effect of the system Hamiltonian
(including dipole-dipole interactions) in addition to external operations.
In every case except for very simple ones, this algorithm must first be 
computed, presumably with the help
of a classical computer. This step in QC may be termed "quantum compilation".
The issue that qubit-qubit interactions raise is
not one of decoherence, but rather whether the determination of the
algorithm, the complation step, becomes prohibitively difficult. This could happen for two
reasons. One is that the interactions are so poorly known that they cannot
be corrected for. It seems likely that quantum error correction can resolve
this difficulty. \ A second and more interesting possibility that the
interactions convert the computation of the algorithm itself into a problem
that grows exponentially with the size of the system. \ We regard this as an
open question and a deep one, that combines many-body theory with algorithm
design and error correction. \ We note that in NMR implementations the
interaction between the qubits also cannot be turned off, but it can be
canceled by refocusing \cite{chuang2}.

In this paper, our aim is to evaluate the importance of spin-phonon coupling
as a source of decoherence in quantum dot qubits. Fundamentally, the issue
is whether the long relaxation times $T_1$ observed at low temperatures in
bulk Si:P carry over to SiGe dots proposed for QC. In Sec. \ref{sec:well} we
introduce the structures that we are interested in. In Sec. \ref%
{sec:decoherence} the physics of the spin-phonon relaxation mechanism is
described qualitatively. In Sec. \ref{sec:si} calculations and results for
pure Si systems are given as a function of the critical design and operating
parameters. In Sec. \ref{sec:ge} are found the correponding results for
structures containing Ge. Sec. \ref{sec:conclusion} is the conclusion.

\section{The strained silicon quantum well}

\label{sec:well}

Si-Ge heterostructures are utilized widely in the digital electronics
industry, and presently have the shortest switching times of any device. One
reason for their success lies in the ability to engineer structures of near
perfect purity, with control over thicknesses and interfaces that approaches
atomic precision -- a technological tour-de-force. An equally key
achievement has been the harnessing of strain as a tool to control band
offsets in heterostructure devices. This paper presents calculations of spin
relaxation for real SiGe structures such as those proposed by Vrijen \textit{%
et al.} \cite{vrijen} and Friesen \textit{et al.} \cite{eriksson}.
Accordingly, we have calculated the electron wavefunctions in quantum wells,
which is needed as input for these calculations. Details of these
calculations are presented in Ref. \cite{eriksson}, and will not be repeated
here. In this section we only describe those aspects of the calculations
that are germane to spin relaxation.

Quantum wells are constructed by sandwiching a very thin layer of one
material between two others. Electrons can be confined in the quantum well
layer when the conduction band offsets produce a potential well. The key to
this technology is therefore to understand the band structure of the various
layers. In this section we will consider a particular class of wells formed
of pure Si, sandwiched between barrier layers of SiGe. \ We will find that
this is optimal from the standpoint of spin coherence. Metallic gates or
impurities create zero-dimensional bound states that define a quantum dot. \
We first review briefly effective mass theory for dots in pure, unstrained
Si, then unstrained SiGe, and finally strained Si.

In pure Si, the $\Delta $ conduction band minima occur near the symmetry
points $X$, in the directions $\{001\}$. In a perfect Si crystal these
minima are six-fold degenerate, the valleys being equivalent. \ In the dot
the electron feels a potential $V_{g}(\vec{r})$ in addition to the atomic
potential, which lifts the degeneracy, though the splittings are not large.
\ The spatial variation of $V_{g}(\vec{r})$ is on length scales generally
much longer than the lattice spacing. For the moment, we shall assume that
the electron is in the ground state of $V_{g}(\vec{r})$ and ignore mixing
with any excited states. \ To the extent that the scale of variation of $%
V_{g}(\vec{r})$ is much longer than the atomic spacing there are six nearly
degenerate ground states. \ This is referred to as the ''valley
degeneracy''. The wavefunctions can be written as\cite{kohn57} 
\begin{equation}
\Phi _{n}(\vec{r})=\sum_{j=1}^{6}\alpha _{n}^{(j)}F_{j}(\vec{r})\phi _{j}(%
\vec{r}).  \label{eq:psi6}
\end{equation}%
Here, $\phi _{j}$ is a Bloch function of the form 
\begin{equation}
\phi _{j}(\vec{r})=u_{j}(\vec{r})e^{i{\vec{k}}_{j}\cdot \vec{r}},
\label{eq:bloch}
\end{equation}%
where $\vec{k}_{j}$ are the six $\Delta $ minima $\{+k_{0}\hat{x},-k_{0}\hat{%
x},+k_{0}\hat{y},-k_{0}\hat{y},+k_{0}\hat{z},-k_{0}\hat{z}\}$ (we shall
always use this ordering), and $u_{j}(\vec{r})$ are periodic functions with
the same periodicity as the crystal potential $V_{p}(\vec{r})$. The $F_{\pm
z}$ are envelope functions that satisfy the Schr\"{o}dinger-like equation 
\begin{equation}
\left[ -\frac{\hbar ^{2}}{2m_{l}}\frac{\partial ^{2}}{\partial z^{2}}-\frac{%
\hbar ^{2}}{2m_{t}}\left( \frac{\partial ^{2}}{\partial x^{2}}+\frac{%
\partial ^{2}}{\partial y^{2}}\right) +V_{g}(\vec{r})\right] F_{\pm z}(\vec{r%
}))=(E-E_{k_{z}}^{(\Delta )})F_{\pm z}(\vec{r}),  \label{eq:envelope}
\end{equation}%
and are independently normalized to unity, similar to wavefunctions.
Analogous equations can be given for the $\pm \hat{\mathbf{x}}$ and $\pm 
\hat{\mathbf{y}}$ minima. We see that $F_{x}=F_{-x}$, $F_{y}=F_{-y}$, and $%
F_{z}=F_{-z}$, so only three independent envelope functions must be
computed. $m_{l}$ and $m_{t}$ are the longitudinal and transverse effective
masses associated with the anisotropic conduction band valleys. $%
E_{k_{z}}^{(\Delta )}$ is the $\Delta $ conduction band edge at $\mathbf{k}%
_{z}$. The splitting of the degeneracy comes from corrections to this
envelope fuction approximation. \ Different choices of the constants $\alpha
_{n}^{(j)}$ determine the six states. \ \ Their values will be discussed in
Sec. \ref{sec:decoherence}. \ This formalism is a good approximation for
both dot and impurity bound states, as the valley splitings are much smaller
than the energy scales in \ref{eq:envelope}.

Germanium is completely miscible in Si, forming a random alloy. For variable
Ge content $x$, Si$_{1-x}$Ge$_{x}$ exhibits materials properties that vary
gradually over the composition range. The alloy lattice constant, $a_{0}(x)$%
, follows a linear interpolation between pure Si and Ge, known as Vegard's
law, quite accurately for all $x$: $a_{0}(x)=(1-x)a_{\mathrm{Si}}+xa_{%
\mathrm{Ge}}$ \cite{windl98}. \ Electronic properties show an abrupt change
in behavior near $x\simeq 0.85$, where the Si-like $\Delta $ minima cross
over to four-fold degenerate, Ge-like, $L$ minima. In this work we focus on
the range $x\lesssim 0.5$, which is strictly Si-like, though we will have
some remarks below on Ge-rich structures. Throughout this range, properties
such as effective mass and the dielectric constant vary only slightly from
pure Si values. For our calculations, the most important parameter is the
conduction band edge, $E^{(\Delta )}(x)$, which remains 6-fold degenerate in
the range $x\lesssim 0.5$. The theory of the variation of $E^{(\Delta )}$%
with $x$ is not germane to the present work, and we simply quote the
empirical result, linear in $x$, which is consistent with Ref. \cite%
{rieger93}: 
\begin{equation}
\Delta E^{(\Delta )}(x)=E^{(\Delta )}(x)-E^{(\Delta )}(0)\simeq 0.23x(%
\mathrm{eV}).  \label{eq:unstrained}
\end{equation}%
(We note, however, that Ref. \cite{schaeffler97} suggests a slope for $%
E^{(\Delta )}(x)$ of opposite sign.) The relatively weak variations of the
effective mass and the dielectric constant will be ignored here.

We consider thin Si wells in which the Si layer grows pseudomorphically. The
in-plane lattice constant, $a_{\Vert }$, must be the same for all the
layers, causing a tetragonal distortion in the strained layer(s). Here we
consider the case of strained Si grown on the (001) surface of
relaxed Si$_{1-x}$Ge$_{x}$.
The in-plane Si lattice constant depends on $x$ as 
\begin{equation}
a_{\Vert }(x)=(1-x)a_{\mathrm{Si}}+xa_{\mathrm{Ge}}.  \label{eq:vergard}
\end{equation}%
Since $a_{\mathrm{Ge}}>a_{\mathrm{Si}}$, the Si is under tensile strain
in the plane.
Hence, the out-of-plane Si lattice constant, $a_{\perp }$, is reduced
according to continuum elastic theory: 
\begin{equation}
a_{\perp }(x)=a_{\mathrm{Si}}\left[ 1-2\frac{c_{12}}{c_{11}}\frac{a_{\Vert
}(x)-a_{\mathrm{Si}}}{a_{\mathrm{Si}}}\right] ,  \label{eq:aperp}
\end{equation}%
where $c_{11}$ and $c_{12}$ are elastic constants for pure Si.

Strain produces shifts of the $\Delta $ band proportional to the strain
variables 
\begin{equation}
\varepsilon _{\Vert }(x)=\frac{a_{\Vert }(x)-a_{\mathrm{Si}}}{a_{\mathrm{Si}}%
}\quad \quad \mathrm{and}\quad \quad \varepsilon _{\perp }(x)=\frac{a_{\perp
}(x)-a_{\mathrm{Si}}}{a_{\mathrm{Si}}}.  \label{eq:xi}
\end{equation}%
with proportionality constants called the dilational and uniaxial
deformation potentials, $\Xi _{d}^{(\Delta )}$ and $\Xi _{u}^{(\Delta )}$,
respectively. Because of the anisotropic nature of the strain, the two $\hat{%
\mathbf{z}}$ minima are shifted down relative to the $\hat{\mathbf{x}}$ and $%
\hat{\mathbf{y}}$ minima, resulting in a splitting of the $\Delta $
conduction band. The net shifts with respect to the unstrained Si $\Delta $
band are given by \cite{rieger93} 
\begin{eqnarray}
\Delta E^{(\Delta _{\perp })}(x) &=&(\Xi _{d}^{(\Delta )}+\frac{1}{3}\Xi
_{u}^{(\Delta )})(2\varepsilon _{\Vert }(x)+\varepsilon _{\perp }(x))+\frac{2%
}{3}\Xi _{u}^{(\Delta )}(\varepsilon _{\perp }(x)-\varepsilon _{\Vert }(x)),
\label{eq:dEperp} \\
\Delta E^{(\Delta _{\Vert })}(x) &=&(\Xi _{d}^{(\Delta )}+\frac{1}{3}\Xi
_{u}^{(\Delta )})(2\varepsilon _{\Vert }(x)+\varepsilon _{\perp }(x))-\frac{1%
}{3}\Xi _{u}^{(\Delta )}(\varepsilon _{\perp }(x)-\varepsilon _{\Vert }(x)).
\label{eq:dEpar}
\end{eqnarray}%
The first terms in Eqs.~(\ref{eq:dEperp}) and (\ref{eq:dEpar}) are
hydrostatic strain terms, which lower the conduction edge compared to
unstrained Si. The second terms in Eqs.~(\ref{eq:dEperp}) and (\ref{eq:dEpar}%
) produce the splitting, associated with uniaxial strain. To perform our
calculations, we use the materials parameters given in Table~\ref{tab:params}%
. However we note that the deformation potentials, particularly $\Xi
_{d}^{(\Delta )}$, are very difficult to measure experimentally.
Considerable disagreement exists in the literature as to the value and even
the sign of $\Xi _{d}^{(\Delta )}$ \cite{fischetti96}. The value given in
Table~\ref{tab:params} was reported (but not endorsed!) in Ref. \cite%
{fischetti96}, and provides energy band variations in general agreement with
Refs.~\cite{rieger93} and \cite{schaeffler97}. We arrive at the following
strain-induced shifts of the conduction band edge for pure Si: 
\begin{equation}
\Delta E^{(\Delta _{\perp })}=-0.86x\quad \quad \mathrm{and}\quad \quad
\Delta E^{(\Delta _{\Vert })}=-0.16x.  \label{eq:strained}
\end{equation}%
The corresponding shift in the relaxed barrier layers, due to the presence
of Ge, was given in Eq.~(\ref{eq:unstrained}). Together, these results
describe the conduction band offsets for the quantum well that are used in
our simulations.

\begin{figure}[ht]
\includegraphics{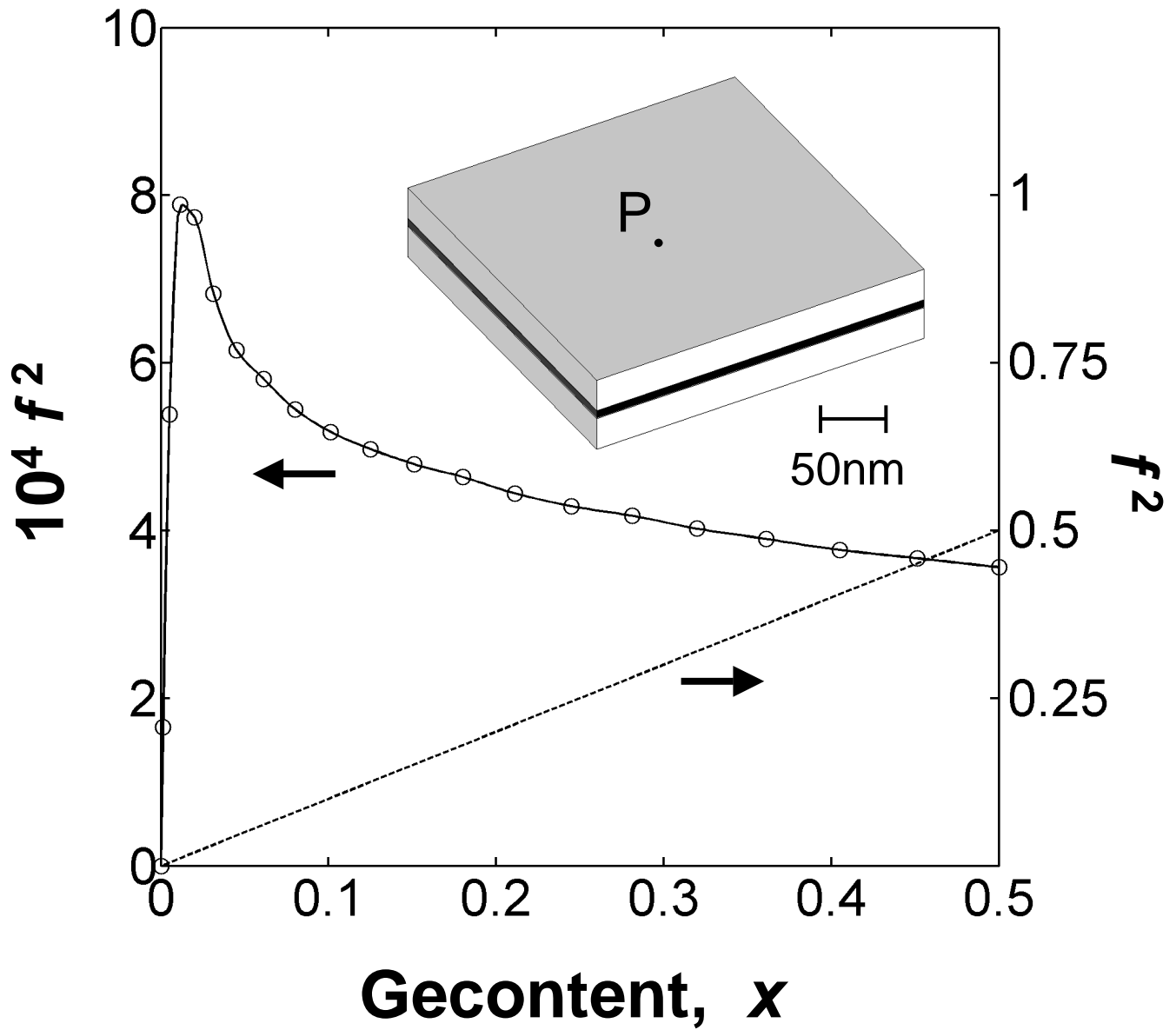}
\caption{Probability, $f^2$, for finding a donor-bound electron at a Ge
site, as a function of Ge content, $x$. The simulated structure, design (1),
is shown in the inset. A strained Si quantum well of thickness 6 nm is
sandwiched between relaxed Si$_{1-x}$Ge$_x$ barrier regions of thickness 20
nm. The electron is bound to a P$^{1+}$ ion embedded in the center of the
quantum well. }
\label{fig:donor}
\end{figure}

\begin{figure}[ht]
\includegraphics{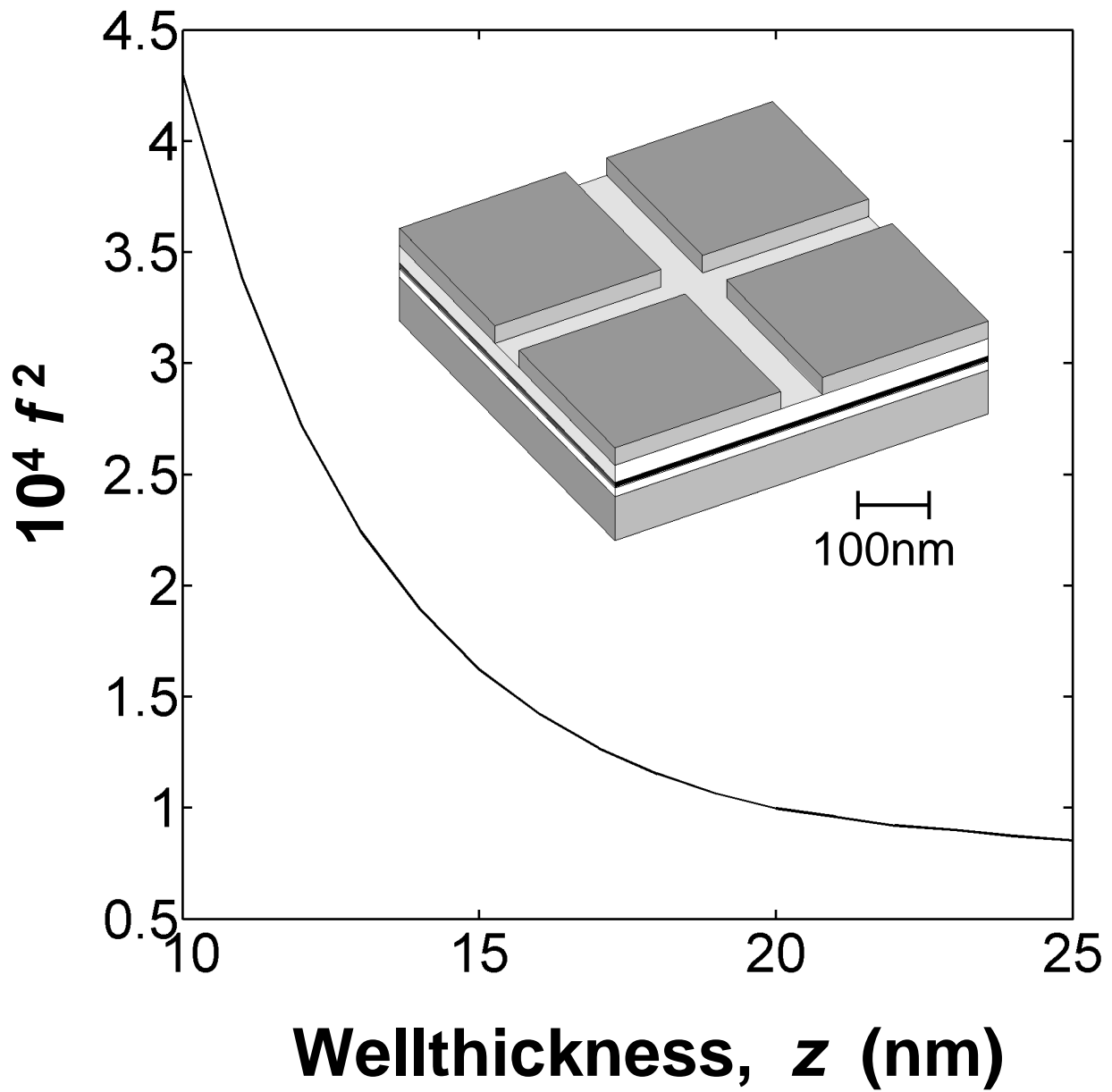}
\caption{Probability, $f^2$, for finding an electrostatically bound electron
at a Ge site, as a function of quantum well thickness, $z$. The inset shows
the heterostructure layers for design (2), beginning at bottom: a thick,
doped semiconductor back-gate, a relaxed Si$_{1-x}$Ge$_x$ barrier layer, a
strained Si quantum well, a thick, relaxed Si$_{1-x}$Ge$_x$ barrier layer,
and lithographically patterned metallic top-gates. The distance between
back- and top-gates is held fixed at 40 nm, while the quantum well, of
variable thickness, $z$, is centered 15 nm above the back-gate.}
\label{fig:qw}
\end{figure}

We now apply our results to two specific quantum well designs of interest
for quantum computing. Design (1), shown in the inset of Fig.~\ref{fig:donor}%
, is a version of that proposed by Vrijen \textit{et al.}\ \cite{vrijen}, in
which electrons are trapped on donor ions (usually P), implanted in a
semiconductor matrix. In that work, the quantum well is split into Ge-rich
and Si-rich regions to facilitate single qubit operations. For simplicity,
we consider here a uniform quantum well, formed of pure Si, with a single
dopant ion located at the center of the well. In such a device, single qubit
operations can be accomplished using a coded qubit scheme \cite{divincenzo00}%
. Design (2), proposed by Friesen et al. \cite{eriksson}, is shown in Fig.~%
\ref{fig:qw}. The confinement potential for the electrons is much softer
than in design (1). Electrons are trapped vertically by the quantum well,
and laterally by the electrostatic potential arising from lithographically
patterned, metallic top-gates. Additionally, the quantum dot is
tunnel-coupled to a degenerate doped back-gate. The dimensions for both
designs are given in the figures.

The wave function of the bound electron is computed in the envelope function
formalism, Eq. (\ref{eq:envelope}). Couplings between the different valleys
are introduced through the perturbation theory described in Sec.~\ref{sec:si}%
. This procedure provides the specific values of $\alpha _{g}^{(i)}$ for the
ground state, which we use in our calculations.

The abrupt conduction band offsets are handled by matching the ground state
wave function $\Phi_g(\vec {r})$ and $\partial_{z}\Phi_g(\vec {r})$ at the
interfaces. (Remember that we have equated effective masses on both sides of
interfaces.) Due to linear independence of the Bloch functions, the boundary
conditions do not cause mixing of the envelope functions. Solutions of Eq.~(%
\ref{eq:envelope}) and the analogous $F_{x}$ equation are obtained, using
commercial 3D finite element software.

As will be seen in Sec. \ref{sec:ge}, the key quantity for the
computation of $T_1$ is $f^{2}$, which describes the probability for the
bound electron to be on a Ge atom. Ge is associated with reduced coherence
times, by virtue of its large spin-orbit coupling. Referring to Eq. (1),
one deduces that this probability may be expressed as 
\begin{equation}
f^{2}=x\left[ 4(\alpha_g^{(x)})^{2}\int_{\Omega _{b}}d^{3}r\,F_{x}^{2}(%
\mathbf{r})+2 (\alpha_g^{(z)})^{2}\int_{\Omega _{b}}d^{3}r\,F_{z}^{2}(%
\mathbf{r})\right] ,
\end{equation}%
where $\Omega _{b}$ is the volume outside the quantum well, if the well is
pure Si. The subscript $g$ refers to the ground state.
The term in square brackets reflects the probability of
finding the bound electron in a barrier region, while $x$ gives the
probability that the electron is on a Ge site.

Fig.~\ref{fig:donor} shows the results of our calculations for $f^{2}$ in
design (1), as a function of the Ge concentration, $x$, in the Si$_{1-x}$Ge$%
_{x}$ barriers. For $x>0.02$, $f^{2}$ decreases with $x$ for two reasons.
First, as $x$ increases, the conduction band offset at the quantum well also
increases, allowing less of the wavefunction to penetrate the barrier.
Second, the spatial extent the electron in the $\hat{\mathbf{z}}$ direction
is greater for $F_{x}$ than $F_{z}$, because of the anisotropic effective
mass . However, less of $F_{x}$ is mixed into the wavefunction for large $x$%
, since $\alpha _{x}$ becomes very small. For $x\lesssim 0.02$, $f^{2}$
drops quickly to zero, due to the absence of Ge in the barriers. \ In the
actual design of Vrijen \textit{et al.}, \cite{vrijen} there is Ge in the
active layer. \ To give an idea of the effect of this, we include an
equivalent value of $f^{2}$ for such a structure.

Fig.~\ref{fig:qw} shows results of the $f^{2}$ calculation for design (2),
as a function of the quantum well thickness, $z$. To perform the
calculations, we have considered a fixed Ge concentration, $x=0.05$, and
taken the limit of large strain, so that $\alpha _{g}^{(\pm x)}\simeq \alpha
_{g}^{(\pm y)}\simeq 0$ and $\alpha _{g}^{(\pm z)}\simeq 1/\sqrt{2}$ for the
ground state. As $z$ increases, less of the wavefunction penetrates the
barrier regions, causing $f^{2}$ to decrease.

\section{Spin relaxation due to coupling to phonons}

\label{sec:decoherence}

In this section we give the method for calculating $T_{1}$, the spin-flip
time of a spin qubit in the ground orbital state due to emission or
absorption of a phonon, following the logic used by Hasegawa \cite{hasegawa}
and Roth \cite{roth} for bulk Si.

Consider a single impurity with a unit positive charge, such as a phosphorus
atom, at the origin. \ In the absence of central cell corrections, there is
a 12-fold degenerate ground state, including spin. \ This valley degeneracy
of the ground state is reduced to two by these corrections, and the
splitting between the 2-fold spin-degenerate ground state and the higher
states is of order $\Delta E\sim 10\,meV.$ \ We shall discuss the detailed
linear combinations ($\alpha $-values) of the states below, as the
coefficients giving the various valley amplitudes play an important role in
the calculation of matrix elements. \ These twelve states may all be thought
of as hydrogenic $1s$ states. \ The splitting of $1s$ and $2s$ is about $%
30~meV$, larger than the $10~meV$ valley splittings. Let us now split the
2-fold degenerate ground state by applying a DC magnetic field in the $z$%
-direction. \ The transition rates between these states are denoted by $%
W_{\uparrow \downarrow }$ and $W_{\downarrow \uparrow }$. The relaxation
time $T_{1}$ is defined by $1/T_{1}=$ \ $W_{\uparrow \downarrow
}+W_{\downarrow \uparrow }$.

The transitions are caused by phonons, but there are important approximate
symmetries that suppress these transitions. \ These are: \ (1) spin rotation
symmetry, meaning that the electron spin cannot be flipped by a phonon; this
symmetry is broken by SOC; (2) time-reversal symmetry, meaning that one
state cannot be changed into its time-reversed partner by emission or
absorption of a phonon; this symmetry is broken by the external magnetic
field; (3) point group symmetries; these are partially broken when strain is
applied.

The spin rotation symmetry would rule out phonon-mediated transitions
between the two states entirely if there were no spin-orbit coupling (SOC).
\ This means that the effects of SOC on the wavefunctions, even though these
effects are small in relatively low-$Z$ Si, must be taken into account. \
When we refer to a state as $\uparrow $ or $\downarrow $, these symbols must
be taken to refer to the majority spin content of the state, not to a pure
spin state. \ Transition rates are roughly proportional to $(g-2)^{2}$,
[more precisely $(g_{l}-g_{t})^{2}$, where $g_{t}(g_{l})$ is the transverse
(longitudinal) $g$-factor, see below for definitions]. \ 

The time-reversal symmetry implies that transitions cannot take place
directly between Kramers-degenerate states even in the presence of
spin-orbit coupling. \ The direct phonon-mediated transitions between the
two states of interest to us are strongly suppressed by this approximate
symmetry. \ It is broken only by the external field $H$. \ The fastest
processes then involve virtual excitation to higher-energy states that are
mixed into the ground state by $H$. \ Hence $1/T_{1}$ involves a factor $%
\left( \mu _{B}H/\Delta E\right) ^{2}$. \ There is an additional factor of $%
H^{2}$ from the phonon density of states, giving an overall rate $%
1/T_{1}\sim H^{4}$ in the limit of small $H$.

The point group symmetry is reduced from cubic to tetragonal under strain. \
This has complex effects that we will explain below.

Before giving actual calculations, we summarize those differences between
the electrons in donor impurity states and in an artificial dot that affect $%
T_{1}$. The most obvious is the single-particle potential that binds the
electron. \ The gate potential is much smoother than the hydrogenic
potential of the impurity. \ This implies that the corrections to the
effective mass approximation are much weaker, and $\Delta E$ will be much
reduced. \ It is difficult to compute the energy splittings precisely, but
considerations based on the method of Sham and Nakayama \cite{sham79} give
splittings in the range $\sim 0.05-0.1~meV$ in the structure of Friesen et
al. \cite{eriksson}. \ This increases the relaxation rate. \ (In fact a
naive estimate of the enhancement is a factor of 400.) On the other hand,
the structures we consider have strong lattice strain. \ This partly lifts
the valley degeneracy and also reduces the matrix elements, which decreases
the rate. \ Another aspect of some of the proposed designs is the presence
of Ge with its much stronger SOC. \ This will act to decrease the spin
relaxation time.

\section{Pure Si quantum dots}

\label{sec:si} We consider first the case of pure Si under uniaxial strain.
\ The ingredients of the calculation are as follows. \ 

From Sec. \ref{sec:well} we have the solutions to the Schr\"{o}dinger
equation $[\mathcal{H}_{0}+ V_{g}(\vec{r})]\Phi_{n0}(\vec{r}) =
E_{n}\Phi_{n0}(\vec{r})$. $\mathcal{H}_{0}$ is the unperturbed crystal
Hamiltonian without SOC and it has the full space group symmetry. $V_{g}(%
\vec{r})$ is the gate and/or impurity potential.

To calculate $T_{1}$, we must also include SOC, which we treat as a
perturbation: $H_{SOC}=\lambda _{Si}\sum_{\vec{R}}\vec{L}_{\vec{R}}\cdot 
\vec{S}_{\vec{R}}$. The resulting states $\Phi _{n}(\vec{r})$ are two-fold
degenerate because of time-reversal symmetry. \ Let us denote these states
as $\Phi _{n\uparrow }(\vec{r})$ and $\Phi _{n\downarrow }(\vec{r})$. They
are not eigenstates of spin, so the arrows denote pseudo-spin. \ We may
define the pseudo-up state as that which evolves from the spin-up state as
spin-orbit coupling is turned on adiabatically, and similarly for the
pseudo-down state. Because of valley and pseudospin degeneracy, there are 2
ground states $\Phi _{gs}(\vec{r})$ and 10 excited states $\Phi _{rs}(\vec{r}%
).$ \ The 12-fold degeneracy in the effective mass approximation is broken
by central-cell corrections in the impurity case and smaller corrections in
the quantum dot.

There is also the Zeeman Hamiltonian of the external field $\mathcal{H}%
_{Z}=\mu _{B}\vec{B}\cdot (\vec{L}+2\vec{S}).$ \ In the field $\Phi
_{n\uparrow }(\vec{r})$ and $\Phi _{n\downarrow }(\vec{r})$ are no longer
degenerate. \ Note that the energy splitting may depend on the direction of $%
\vec{B}$.

Finally, we have the electron-phonon coupling Hamiltonian $\mathcal{H}_{ep}.$
\ A phonon represents a time-dependent perturbation. \ This will create
transitions whose rate is given by the Fermi golden rule. \ We are
interested in the transitions between $\Phi _{g\uparrow }(\vec{r})$ and $%
\Phi _{g\downarrow }(\vec{r}).$ \ However, $\left\langle \Phi _{g\uparrow }(%
\vec{r})|\mathcal{H}_{ep}|\Phi _{g\downarrow }(\vec{r})\right\rangle =0$ in
\ the absence of the external field. \ Thus we need to calculate in next
order in perturbation theory using an effective Hamiltonian%
\begin{equation}
\mathcal{H}^{\prime }=\sum_{rs}\frac{1}{E_{g}-E_{r}}\left\{ \left[ \left. 
\mathcal{H}_{Z}|\Phi _{rs}(\vec{r})\right\rangle \left\langle \Phi _{rs}(%
\vec{r})|\mathcal{H}_{ep}\right. \right] +\left[ \left. \mathcal{H}%
_{ep}|\Phi _{rs}(\vec{r})\right\rangle \left\langle \Phi _{rs}(\vec{r})|%
\mathcal{H}_{Z}\right. \right] \right\}
\end{equation}%
Here $r$ runs over the excited states, $r=2,..,6$. \ $s=\uparrow ,\downarrow 
$. \ 

The relaxation time is given by $1/T_{1}=W_{\uparrow \downarrow
}+W_{\downarrow \uparrow },$ where $W_{\downarrow \uparrow }=(2\pi /\hbar
)\times \sum_{\vec{q}\lambda }\left| \,\left\langle \Phi _{g\uparrow }(\vec{r%
})|\,\mathcal{H}^{\prime }\,|\Phi _{g\downarrow }(\vec{r})\right\rangle
\right| ^{2}\delta (E_{g\downarrow }-E_{g\uparrow }-\hbar \omega _{\vec{q}%
\lambda })$ $\ [1+n(\omega _{\vec{q}\lambda })]$ is the rate for transitions
from the higher-energy (pseudo-spin down) state to the lower-energy
(pseudo-spin up) state$\ $and $W_{\uparrow \downarrow }=(2\pi /\hbar )\sum_{%
\vec{q}\lambda }\left| \,\left\langle \,\Phi _{g\downarrow }(\vec{r})\,|%
\mathcal{H}^{\prime }|\,\Phi _{g\uparrow }(\vec{r})\right\rangle \right|
^{2}\,\delta (E_{g\downarrow }-E_{g\uparrow }-\hbar \omega _{\vec{q}\lambda
})\,n(\omega _{\vec{q}\lambda })$ is the rate for transitions from the
lower-energy (pseudo-spin up) state to the higher-energy (pseudo-spin down)
state$\ $where the sum is over phonon modes $\vec{q}\lambda $ with energies $%
\omega _{\vec{q}\lambda }$ . \ A thermodynamic average over the lattice
states has been taken. \ It yields the Bose occupation factors $n(\omega _{%
\vec{q}\lambda })$ for the phonons.

The matrix elements of $\mathcal{H}^{\prime }$ are computed as follows.

The expectation value of the external field $H_{Z}$ can be written as%
\begin{eqnarray}
\left\langle \Phi _{ns}(\vec{r})|H_{Z}|\Phi _{n^{\prime }s^{\prime }}(\vec{r}%
)\right\rangle &=&\sum_{i=1}^{6}\alpha _{n}^{(i)}\sum_{j=1}^{6}\alpha
_{n^{\prime }}^{(j)}\mu _{B}\vec{B}\cdot g^{(i)}\cdot \vec{\sigma}%
_{s,s^{\prime }}\,\,\delta _{ij}  \nonumber \\
&=&\mu _{B}\vec{B}\cdot \left[ \sum_{i=1}^{6}\alpha _{n}^{(i)}\alpha
_{n^{\prime }}^{(i)}g^{(i)}\right] \cdot \vec{\sigma}_{s,s^{\prime }} 
\nonumber \\
&\equiv &\mu _{B}\vec{B}\cdot \mathbf{D}_{nn^{\prime }}\cdot \vec{\sigma}%
_{s,s^{\prime }}\,\,  \label{eq:ddef}
\end{eqnarray}%
and the tensor $g^{(i)}$ is the effective $g$ factor at the $i$th valley. \
This equation defines the tensor $\mathbf{D}_{nn^{\prime }}$ that
characterizes the coupling of the various states by the external field. \
The principal axes of the $g$ tensor are the same as that of the effective
mass tensor at the $i$th valley. \ It has the form%
\begin{equation}
g^{(\pm x)}=\left( 
\begin{array}{ccc}
g_{l} & 0 & 0 \\ 
0 & g_{t} & 0 \\ 
0 & 0 & g_{t}%
\end{array}%
\right) ,\,g^{(\pm y)}=\left( 
\begin{array}{ccc}
g_{t} & 0 & 0 \\ 
0 & g_{l} & 0 \\ 
0 & 0 & g_{t}%
\end{array}%
\right) ,g^{(\pm z)}=\left( 
\begin{array}{ccc}
g_{t} & 0 & 0 \\ 
0 & g_{t} & 0 \\ 
0 & 0 & g_{l}%
\end{array}%
\right) .
\end{equation}%
\ There are only two independent constants. \ 

A simple example of the diagonal part of the \textbf{D} tensor is that for
the ground state of an impurity in the \textit{unstrained} lattice when the
central cell corrections are included. Then we have $\alpha _{g}^{(i)}=1/%
\sqrt{6}$ and $\mathbf{D}_{gg}$ is proportional to the unit matrix:%
\begin{equation}
\left\langle \Phi _{gs}(\vec{r})|H_{z}|\Phi _{gs^{\prime }}(\vec{r}%
)\right\rangle =g_{g}\mu _{B}\vec{B}\cdot \vec{\sigma}_{ss^{\prime }}\,,\,
\end{equation}%
with%
\begin{equation}
g_{g}=\frac{2}{3}g_{t}+\frac{1}{3}g_{l}.
\end{equation}

The matrix elements between ground and excited states have the form 
\begin{equation}
\left\langle \Phi _{gs}(\vec{r})|H_{Z}|\Phi _{rs^{\prime }}(\vec{r}%
)\right\rangle =g^{\prime }\mu _{B}\vec{B}\cdot \mathbf{D}_{gr}\cdot \vec{%
\sigma}_{ss^{\prime }}\,,
\end{equation}%
with 
\begin{equation}
g^{\prime }=\frac{1}{3}(g_{l}-g_{t})
\end{equation}%
and the tensor $\mathbf{D}_{r}$ is defined by%
\begin{equation}
\mathbf{D}_{gr}=3\sum_{i=1}^{6}\alpha _{g}^{(i)}\alpha _{r}^{(i)}\widehat{k}%
^{(i)}\widehat{k}^{(i)}
\end{equation}%
where $\widehat{k}^{(i)}$ is the ''local'' anisotropy axis. \ If the
original $g$ were isotropic, then $g^{\prime }=0$ and there would be no
coupling between different states and no spin relaxation.

\ If the lattice is strained, then the $\alpha $ coefficients become
strain-dependent and the general expression for $\mathbf{D}$\ from Eq. \ref%
{eq:ddef} must be used. Uniaxial strain lifts the degeneracy of the valleys.
\ We include this effect in the Hamiltonian and it determines the proper
combinations of the $\alpha _{n}^{(i)}$ defined in Eq. \ref{eq:psi6}. These
then feed into $\mathbf{D}_{gr}.$ \ As a function of strain the $\alpha
_{n}^{(g)}$ for the ground state cross over from the completely symmetric
combination $\alpha _{g}^{(i)}=1/\sqrt{6}$ to the combination $\alpha
_{g}^{(\pm x)}=\alpha _{(g)}^{\pm y}=0$, $\alpha _{(g)}^{\pm z}=1/\sqrt{2}$
in the limit of large strain.\cite{feher}

The phonons involved are just the acoustic ones, one longitudinal and two
transverse - these are the only ones with low enough energy to play a role
in relaxing the spins. \ The matrix elements of the electron-phonon
interaction are only nonzero within one valley and for a single phonon mode
they are conventionally parametrized as:%
\begin{equation}
\left\langle \Psi _{c\vec{k}s}^{(i)}(\vec{r})|H_{ep}^{(\vec{q}\lambda
)}|\Psi _{c\vec{k}^{\prime }s}^{(i)}(\vec{r})\right\rangle =ib_{\vec{q}%
\lambda }\widehat{e}_{\lambda }(\vec{q})\cdot (\Xi _{d}1+\Xi _{u}\widehat{k}%
^{(i)}\widehat{k}^{(i)})\cdot \vec{q}+~h.c.
\end{equation}%
near the $i$th valley, where $\vec{q}=\vec{k}-\vec{k}^{\prime }$ and $%
\widehat{e}_{\lambda }$ is the polarization vector. $b_{\vec{q}\lambda }$
destroys a phonon with wavevector $\vec{q}$ and polarization $\lambda $.
Once again, we see that the interaction can be characterized by just 2
parameters, in this case $\Xi _{d}$ and $\Xi _{u}$, as already defined in
Eq. \ref{eq:xi}. \ Performing the integration over the envelope function at
wavector $\vec{q}$ now gives%
\begin{eqnarray}
\left\langle \Phi _{gs}(\vec{r})|H_{ep}^{(\vec{q}\lambda )}|\Phi
_{gs^{\prime }}(\vec{r})\right\rangle _{\vec{q}} &=&(\Xi _{d}+\frac{1}{3}\Xi
_{u})A(\vec{q})\left( b_{\vec{q}\lambda }+b_{\vec{q}\lambda }^{\ast }\right)
\delta _{s,s^{\prime }} \\
\left\langle \Phi _{gs}(\vec{r})|H_{ep}^{(\vec{q}\lambda )}|\Phi
_{rs^{\prime }}(\vec{r})\right\rangle _{\vec{q}} &=&\frac{1}{3}\Xi _{u}\,A(%
\vec{q})\left( i\widehat{e}_{\lambda }(\vec{q})\cdot \mathbf{D}_{gr}\cdot 
\vec{q}\,\,b_{\vec{q}\lambda }-i\widehat{e}_{\lambda }^{\ast }(\vec{q})\cdot 
\mathbf{D}_{gr}\cdot \vec{q}\,\,b_{\vec{q}\lambda }^{\ast }\right) \delta
_{s,s^{\prime }}
\end{eqnarray}%
where 
\begin{equation}
A^{(i)}(\vec{q})=\sum_{\vec{k}}F^{(i)\ast }(\vec{k}+\vec{q})F^{(i)}(\vec{k}%
)=\int d^{3}r\,F^{2}(r)e^{i\vec{q}\cdot \vec{r}}.
\end{equation}%
Thus the electron-phonon interaction involves a form factor for the bound
states. \ Since $F$ is normalized, we have $A^{(i)}(\vec{q})\approx 1$ when
the wavelength of the phonon is much longer than the spatial extent $a^{\ast
}$ of the bound state: $qa^{\ast }<<1$. \ The calculations of \ref{sec:well}
indicate that this is the case. \ It is also independent of $(i).$ \ 

In the golden rule calculation, the energy denoninator $\left(
E_{g}-E_{r}\right) ^{-2}$ will suppress contributions from the excited
states of $V_{g}$. \ Thus we will keep only states that are split off from
the ground state by corrections to the effective mass approximation. \ This
approach works very well in Si:P and should be even better for the quantum
dot.

This produces the golden-rule transition rate%
\begin{eqnarray}
W_{\uparrow \downarrow } &=&\frac{2\pi }{\hbar }\left[ \frac{1}{3}\Xi
_{u}g^{\prime }\mu _{B}\right] ^{2}\sum_{\vec{q}\lambda }\,A^{2}(\vec{q}%
)\,\delta (E_{g\uparrow }-E_{g\downarrow }-\hbar \omega _{\vec{q}\lambda
})\left\langle a_{\vec{q}\lambda }a_{\vec{q}\lambda }^{\ast }\right\rangle
\times  \nonumber \\
&&\left| \sum_{r=2}^{6}\frac{\vec{B}\cdot \mathbf{D}_{gr}\cdot \vec{\sigma}%
_{\uparrow \downarrow }\,\,\widehat{e}_{\lambda }(\vec{q})\cdot \mathbf{D}%
_{gr}\cdot \vec{q}\,\delta _{ss^{\prime }}}{E_{g}-E_{r}}\right| ^{2}.
\end{eqnarray}

We approximate the phonon dispersion as $\omega _{\vec{q}\lambda
}=v_{\lambda }q$. \ Setting $A=1$, performing the integral over the
magnitude of $\vec{q}$, and repeating the calculation for $W_{\downarrow
\uparrow }$, we obtain a total spin relaxation time

\begin{eqnarray}
\frac{1}{T_{s}} &=&W_{\uparrow \downarrow}+W_{\downarrow \uparrow} =\frac{1}{%
8\pi ^{2}\rho \hbar ^{4}}\left( \frac{g^{\prime }\mu _{B}\,B\,\Xi _{u}}{3}%
\right) ^{2}[2\,n(g\mu _{B}B)+1]\,g_{g}^{3}\mu _{B}^{3}B^{3}\times  \nonumber
\\
&&\sum_{\lambda =1}^{3}\frac{1}{v_{\lambda }^{5}}\int_{0}^{2\pi }d\phi
^{\prime }\int_{0}^{\pi }\sin \theta ^{\prime }d\theta ^{\prime }\,\left|
\sum_{r=2}^{6}\left[ \vec{B}(\theta ,\phi )\cdot \mathbf{D}_{gr}\cdot \vec{%
\sigma}_{\uparrow \downarrow }\right] \left[ \frac{\,\,\widehat{e}_{\lambda
}(\vec{q})\cdot \mathbf{D}_{gr}\cdot \widehat{q}\,}{E_{g}-E_{r}}(\theta
^{\prime },\phi ^{\prime })\right] \right| ^{2}.
\end{eqnarray}%
Here $(\theta ,\phi )$ are the polar axes of the direction of $\vec{B}$
measured from the $[100]$ direction of the crystal. $\rho$ is the mass
density. This is our basic result, in a form very similar to that given by
Hasegawa \cite{hasegawa}.

\begin{figure}[ht]
\includegraphics{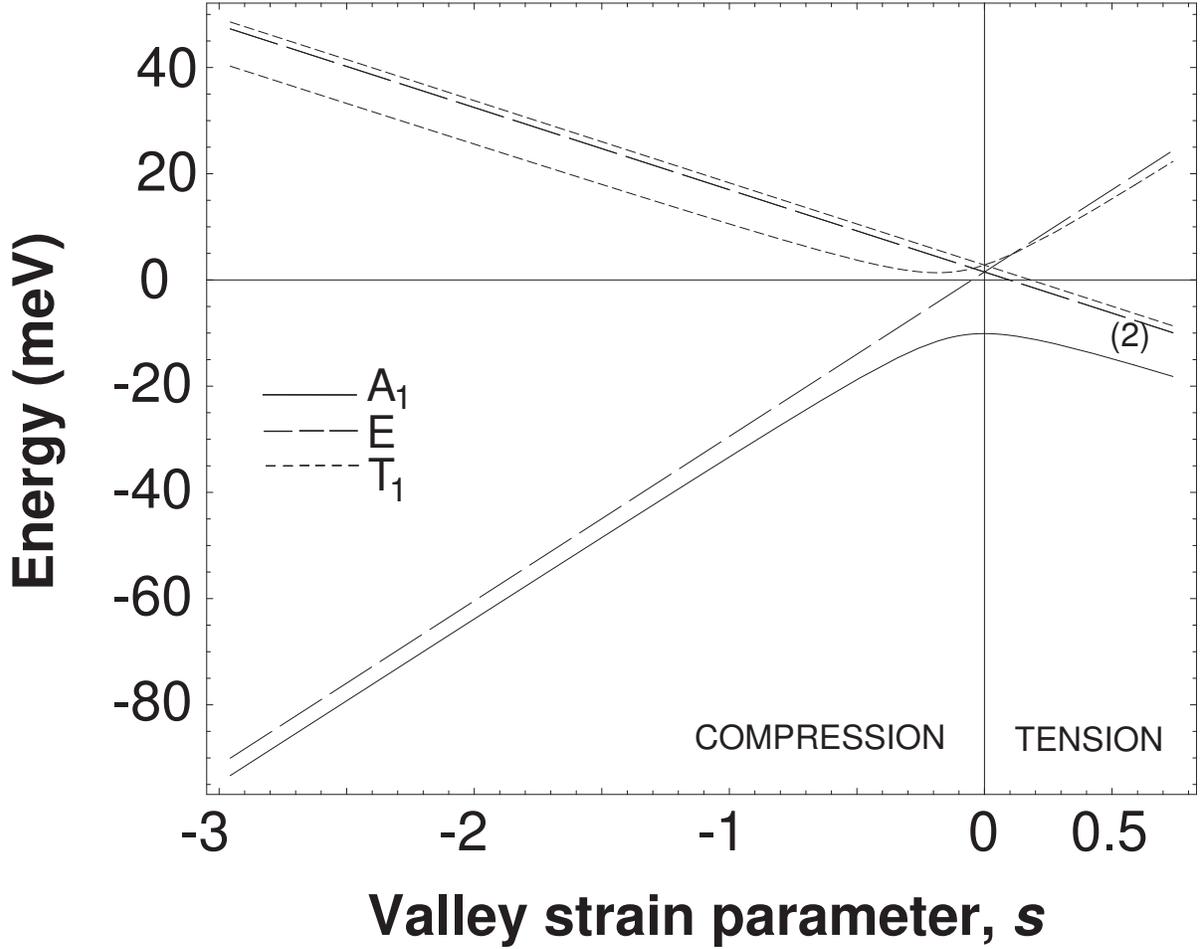}
\caption{Energy of the six 1s-like donor levels of an electron bound by a P
impurity in silicon (with respect to the energy center of gravity) versus
the strain parameter $s$ with an uniaxial stress applied in the [100]
direction. (It is important to note for clarity that what we call compression relative to the [100] or growth direction is equivalent to tension in the plane of the quantum well. This label convention differs across fields.) For reference, $s=-3$ corresponds to the compressive (s negative)
strain caused in a pure silicon layer by a Si$_{0.8}$Ge$_{0.2}$ sublayer.
The energies are expressed in eV and the numbers in parenthesis indicate the
degeneracy of the level. The A$_1$ (ground state, solid line) and T$_1$
level wave functions are mixed by strain (causing valley population
intermixing) and the relaxation rate is proportional to $1/\Delta E$. Thus,
with all else equal, an increase in strain causes the relaxation rate to
decrease. For a detailed analysis see Wilson and Feher \protect\cite{feher},
where $s$ here is their $s^{\prime}$ times 100 or Koiller \textit{et al.} %
\protect\cite{koillerpre} whose notation is $s=100 (6 \protect\chi \Delta_c
/ \Xi_c)$.}
\label{fig:energies}
\end{figure}

In Fig. \ref{fig:energies} the effect of strain on the lowest energy levels
is shown. Here the zero-strain splittings are those of a P impurity. A
similar plot was given in Ref. \cite{feher}, but at that time the correct
zero-strain splittings were not known. \ In the dot, the energy splittings
for the unstrained case would be two to three orders of magnitude smaller
(see below for a discussion).

\begin{figure}[ht]
\includegraphics{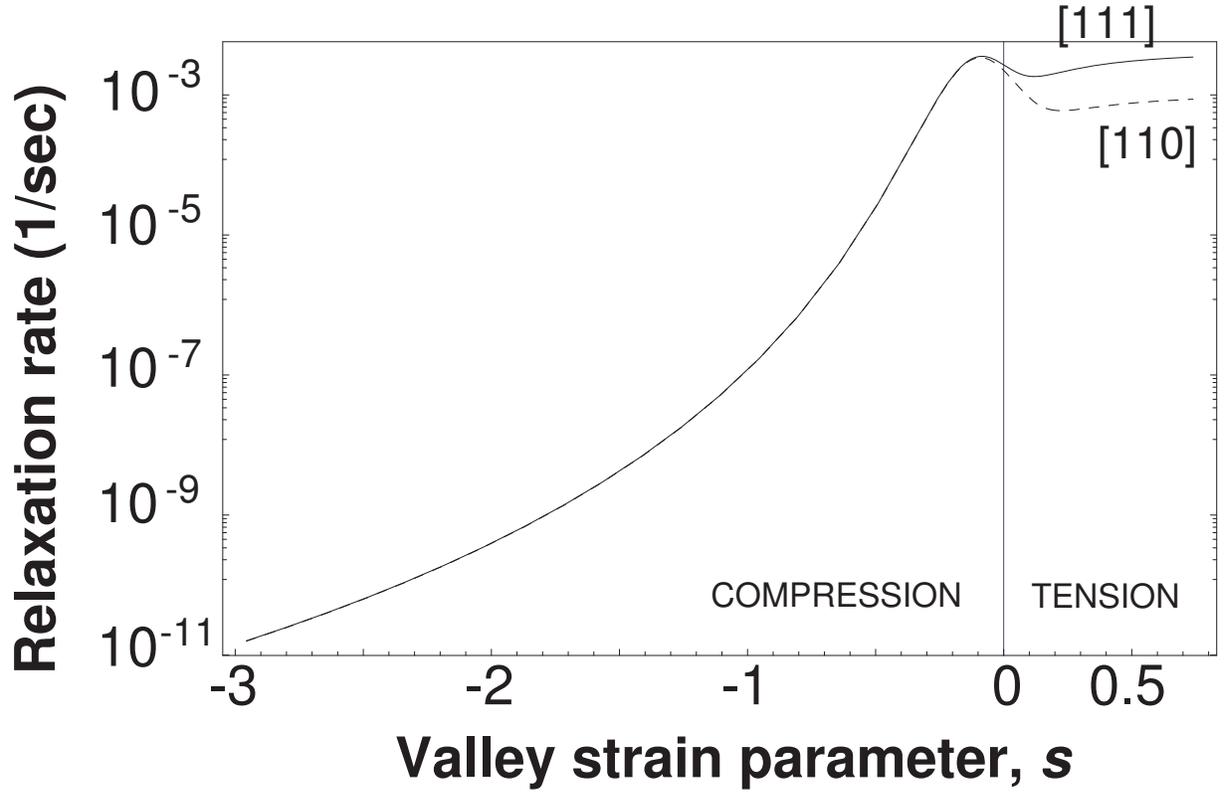}
\caption{Relaxation rates of a P impurity bound electron in [100] uniaxially
strained silicon versus strain parameter $s$ for a temperature of 3 Kelvin
and a magnetic field $H=H(\cos\protect\theta \cos\protect\phi, \cos\protect%
\theta, \sin\protect\theta)$ of 1 Tesla in the [111] ($\protect\theta=%
\protect\phi=\protect\pi/4$) and [110] ($\protect\theta=\protect\pi/4$, $%
\protect\phi=0$) directions respectively. $s=-3$ corresponds to the strain
caused in the pure silicon layer by a Si$_{0.8}$Ge$_{0.2}$ sublayer.}
\label{fig:strain}
\end{figure}

In Fig. \ref{fig:strain}, we show the effect of strain on the relaxation
time of an electron in a P impurity potential. There are two effects: the
overall increase of the energy denominators and the change of the ground
state to a less symmetric valley weighting. This leads to a non-monotonic
dependence of $T_{1}$ on strain, but in the region of interest, the effect
of strain is to greatly increase $T_{1}$, since most proposed structures
have $s<-1.$ \ At large strain, only one energy denominator remains small,
that between the ground state, symmetric in the $\pm z$ valleys, and the
first excited state, antisymmetric in the $\pm z$ valleys. \ The overlap
matrix \textbf{D }is very small between these two state. \ This reduction of
the matrix element is the dominant effect.

\begin{figure}[ht]
\includegraphics{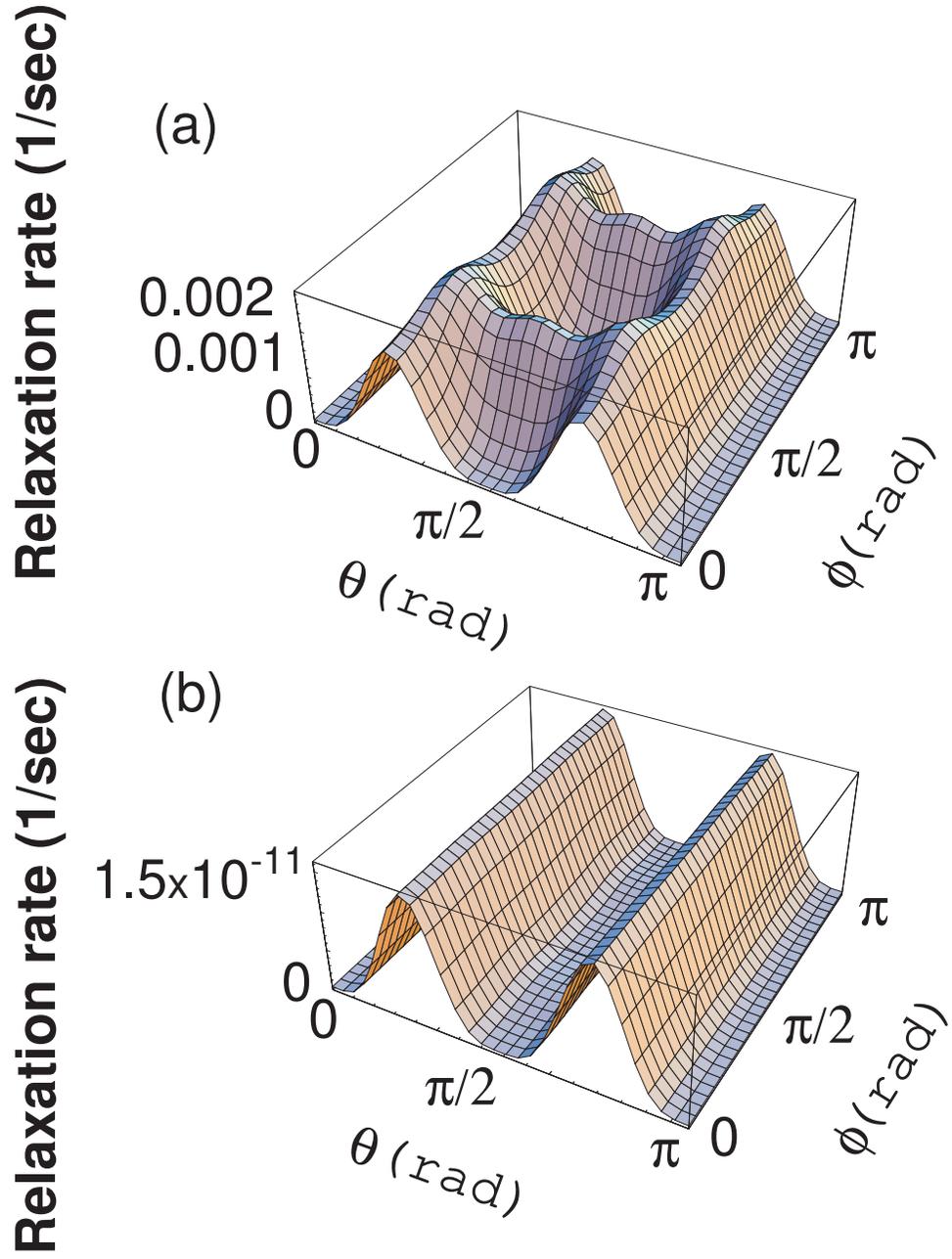}
\caption{Dependence of the relaxation rate on magnetic field direction with $%
H=H(\cos\protect\theta \cos\protect\phi, \cos\protect\theta, \sin\protect%
\theta)$ for (a) unstrained Si and (b) compressively strained Si
(corresponding to a Si$_{0.8}$Ge$_{0.2}$ sublayer). The magnetic field is
set at $1$ T and the temperature is 3 K.}
\label{fig:anisotropy}
\end{figure}

Also of interest is the dependence of $T_{1}$ on the angle of the external
field, since this may serve as a diagnostic tool in experiments to verify
that the relaxation process is really due to spin-phonon coupling. As a
function of strain, this dependence becomes highly anisotropic, as seen in
Fig. \ref{fig:anisotropy}. This is due to the elimination of all but the $%
\pm z$ valleys from the problem at high strain. We note that the limiting
value $T_{1}\rightarrow \infty $ when the external field is along a crystal
axis is cut off by intervalley scattering effects not included in the
present calculation \cite{feher}.

\begin{figure}[ht]
\includegraphics{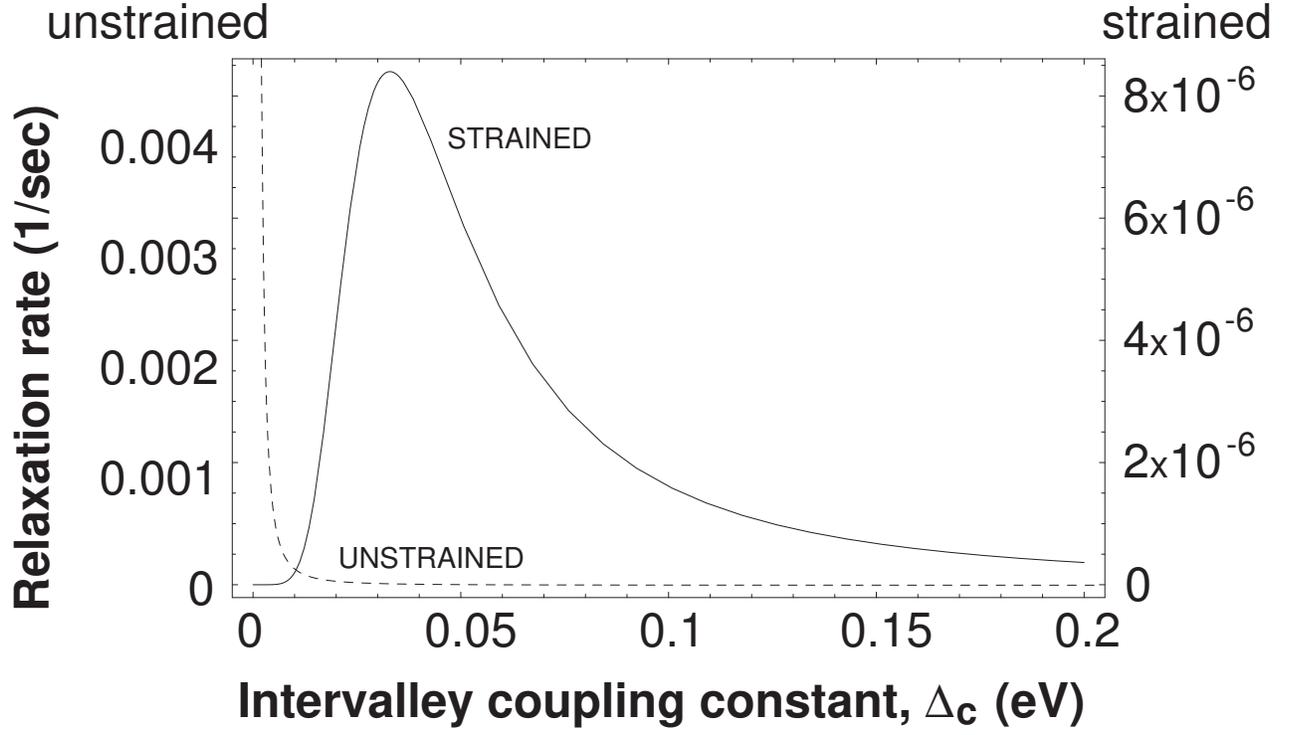}
\caption{Relaxation rate of a bound electron in silicon. The solid line is
for fixed [100] uniaxial strain $s=-3$, corresponding to the strain caused
in the pure silicon layer by a Si$_{0.8}$Ge$_{0.2}$ sublayer. The dashed
line is the unstrained case. The rate is plotted for varying the intervalley
coupling constant $\Delta_c$. $\Delta_c$ corresponds to 1/6th the energy
splitting between the singlet (ground state) and triplet (next excited
state) valley energies {\em in unstrained Si}. As a result, it controls the mixing
between the $\pm x$ and $\pm y$ valley wavefunctions with the $\pm z$ valley
wavefunctions. The magnetic field is set at $1$ T along the [111] direction
and the temperature is 3 K.}
\label{fig:deltac}
\end{figure}

The change in the confining potential reduces the corrections to the
effective mass approximation, as discussed in Sec. \ref{sec:decoherence}
which in turn reduces the energy denominators, leading to a decrease in $%
T_{1}$. In the Friesen \textit{et al.} structure, we have estimated the
splitttings $E_{r}-E_{g}$ using the method of Sham and Nakayama \cite{sham79}%
, and they range from 0.05 to 0.1 meV, depending on the gate voltages that
produce the potential. This may be incorporated into the calculation of the $%
\alpha _{n}^{(i)}$ by introducing a variable coupling $\Delta _{c}$ that
mixes the $\pm x$ and $\pm y$ valley wavefunctions with the $\pm z$ valley
wavefunctions. \ For a precise definition of $\Delta _{c}$, see Ref. \cite%
{feher}. \ Approximately, however, \ $\Delta _{c}\sim
0.2\,(E_{r}-E_{g})<10^{-4}eV$. \ The results of the calculations are shown
in Fig. \ref{fig:deltac}. \ The graphs show the absolutely crucial role that
strain plays in the determination of $T_{1}$. \ Because of the small energy
denominators in the dot, the rate is extremely fast at small $\Delta _{c}$
for the unstrained case: $1/T_{1}\sim \Delta _{c}^{-2}.$ \ 

\begin{figure}[ht]
\includegraphics{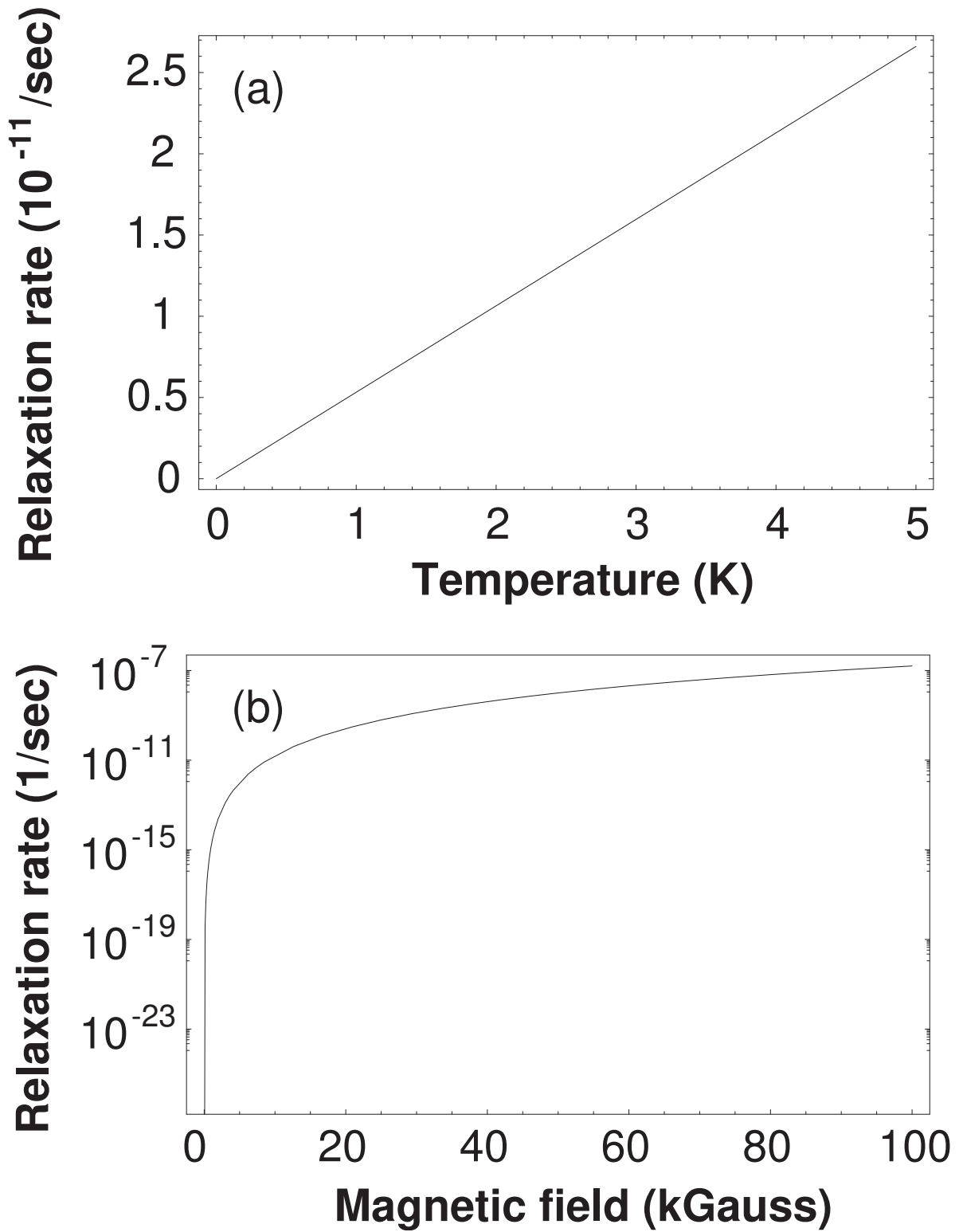}
\caption{Relaxation rates of a P impurity bound electron in [100] uniaxially
strained silicon ($s=-3$, corresponding to the strain caused in the pure
silicon layer by a Si$_{0.8}$Ge$_{0.2}$ sublayer) versus (a) temperature
with the magnetic field set at $1$ T and (b) magnetic field strength with
the temperature set at 3 K.}
\label{fig:temperature}
\end{figure}

We also point out that in stuctures where $T_{1}$ appears to be too long for
efficient preparation of the spins (as discussed in Sec. \ref%
{sec:introduction}) the deficiency can be made up by increasing the
temperature $T$. As seen in Fig. \ref{fig:temperature}, $T_{1}$ decreases
very rapidly as $T$ is increased. Actually, this calculation even
underestimates the decrease, since multiphonon processes begin to contribute
at about 3K. However, it must be borne in mind that the temperature must be
small enough that the initial state is essentially pure (all spins up, for
example). A more practical method of decreasing $T_1$ if desired, would be
to increase the magnetic field.

\section{Structures containing Ge}

\label{sec:ge}

The effect of alloying with Ge is to increase the SOC and hence to increase $%
\left| g-2\right| $. \ First principles calculations of Ge impurities in Si
have shown that there is little effect on the states near the bottom of the
conduction band, though this is not necessarily the case for higher energy
states in the band \cite{jaros}. \ This is in accord with the isoelectronic
character of the atoms in the alloy. \ It is then reasonable to employ the
virtual crystal approximation (VCA). The approximation should be
quantitatively accurate for small, (say $<10\%$) concentrations of Ge, but
may be taken as a good qualitative guide also to higher concentrations. \ \
The Bloch functions in the absence of SOC satisfy%
\begin{equation}
H_{0}\Psi _{n\vec{k}s}=E_{n\vec{k}}\Psi _{n\vec{k}s},
\end{equation}%
where%
\begin{equation}
\Psi _{n\vec{k}s}=\frac{1}{\sqrt{N_{c}}}\sum_{j=1}^{N_{c}}\sum_{l=1}^{2}\exp
(i\vec{k}\cdot r_{jl})\sum_{m=0}^{3}a_{lm}(n\vec{k})\,\phi _{ms}(\vec{r}%
-r_{jl})
\end{equation}%
where $N_{c}$ is the number of unit cells, $j$ labels the unit cells, $l$
labels the two positions in the unit cell, $m$ labels the 4 atomic states, $%
\phi _{ms}$ are the atomic orbitals, and the coefficients $a_{lm}(n\vec{k})$
give the proper linear combination of atomic orbitals (LCAO's) \ for the
state at momentum $\vec{k}$ and band $n$. \ The normalization condition is%
\begin{equation}
\sum_{lm}\left| a_{lm}(n\vec{k})\,\right| ^{2}=1.
\end{equation}

In the VCA, the $a_{lm}(n\vec{k})$ are the same for the Si and Ge sites. \
The SOC Hamiltonian for the alloy is written as

\begin{equation}
H_{SOC}=\sum_{\vec{R}}\lambda_{\vec{R}} \vec{L}_{\vec{R}} \cdot \vec{S}_{%
\vec{R}},
\end{equation}%
where $\lambda _{\vec{R}}$ is the SOC strength for Si (Ge) when $\vec{R}$ is
a Si (Ge) site. \ The energy shift of an electron in the external field $\vec{B} = (B_x , B_y , B_z)$ can
be written as%
\begin{equation}
\Delta E_{n\vec{k}s}(\vec{B})= -\mu _{B}\sum_{i=x,y,z}g_{i}(n\vec{k})B_{i}s,
\end{equation}%
where%
\begin{equation}
g_{i}(n\vec{k})=2-2\hbar ^{2}\left( \sum_{\vec{R}}\lambda _{\vec{R}%
}/N_{c}\right) \sum_{n^{\prime }\neq n,l,m,m^{\prime }}\left| a_{lm}(n\vec{k}%
)\right| ^{2}\left| a_{lm^{\prime }}(n^{\prime }\vec{k})\right| ^{2}\left(
E_{n\vec{k}}-E_{n^{\prime }\vec{k}}\right) ^{-1}(\varepsilon_{imm'})^2 
\nonumber
\end{equation}
and $\varepsilon_{imm'}$ is the completely antisymmetric symbol with $\varepsilon_{123}=1$.

Our approximation now consists in regarding the dependence on the Ge
concentration as coming only in the term $\left( \sum_{\vec{R}}\lambda _{%
\vec{R}}/N_{c}\right) $. \ Hence any component of the $g$-tensor is
proportional to a weighted average of $\lambda _{Si}$ and $\lambda _{Ge}.$ \
In particular, consider the conduction band minimum at $\vec{k}=(0,0,k_{0})$
in the compound Si$_{1-x}$Ge$_{x}.$\ Then we have $g_{z}(x)-2=[g_{z}(0)-2]%
\left[ (1-x)+x(\lambda _{Ge}/\lambda _{Si})\right] ,$ where $g_{z}(0)$ is
the value for pure silicon. \ The numbers, known from atomic physics, are $%
\lambda _{Ge}/\lambda _{Si}=0.295/0.044=6.71$, so $g_{z}(x)-2=\left[
g_{z}(0)-2\right] \left[ (1-x)+6.71x\right] .$ \ Similarly, we have $%
g_{y}(x)-2=\left[ g_{y}(0)-2\right] \left[ (1-x)+6.71x\right] $ , etc. \
Only the overall scale of $g$ changes, not the anisotropy in the VCA. \ 

The presence of Ge in the lattice breaks the translation symmetry, an effect
that is neglected in the VCA. \ Taking this into account would lead to a $g$%
-factor that is averaged over momentum space rather than one that is
evaluated only at the Si conduction band minima. \ If we take the L-minimum
of pure Ge as an example, we would expect that the corrections to $\left|
g-2\right| $ would be greater if there is some averaging over momentum
space. \ Thus a calculation going beyond the VCA would give corrections that
would reduce $T_{1}$. \ 

\begin{figure}[ht]
\includegraphics{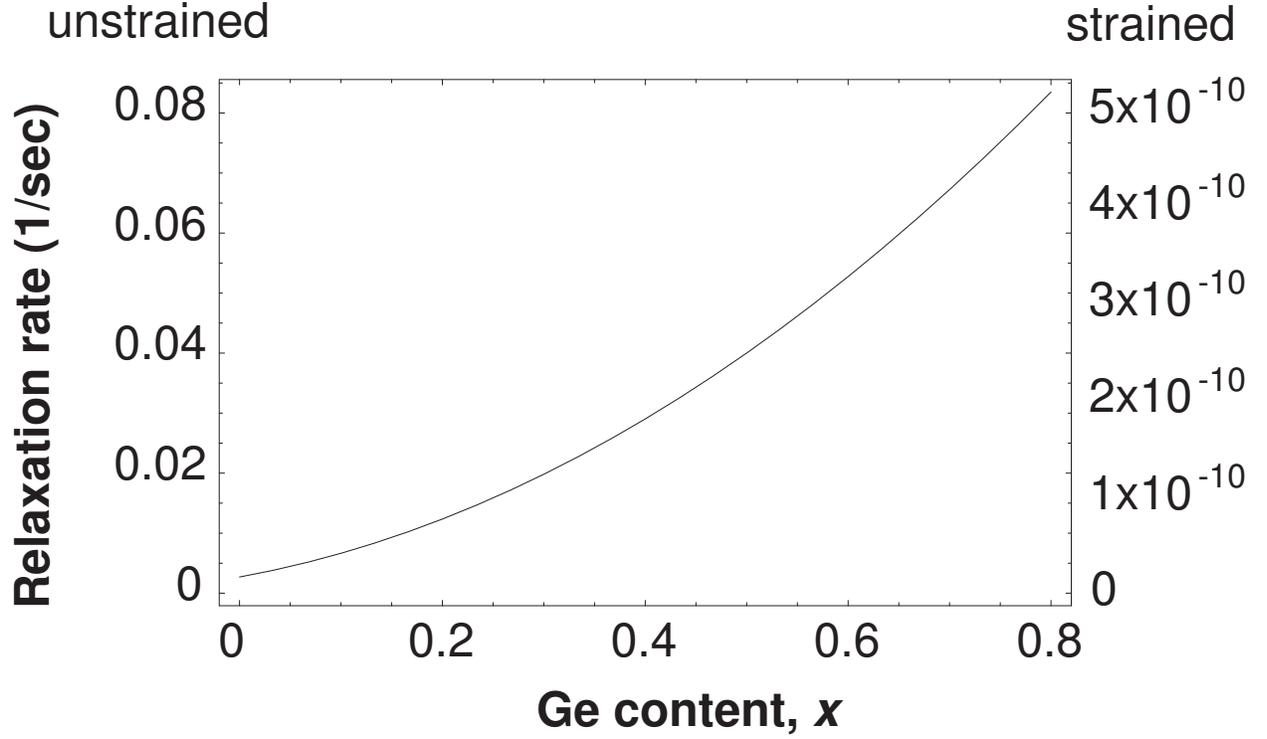}
\caption{Relaxation rate of a P impurity bound electron in [100] uniaxially
strained silicon versus concentration of germanium within unstrained (left
vertical axis) and compressively strained (right vertical axis,
corresponding to a Si$_{0.8}$Ge$_{0.2}$ sublayer) pure silicon. The magnetic
field is set at $1$ T along the [111] direction and the temperature is 3 K.
Here we assume that the addition of germanium does not affect the strain
within the bulk and only acts to increase the $g$-factor and spin orbit
coupling. We expect this approximation to hold for small concentrations of
germanium and to break down with increasing concentration as the conduction
band valley minima switch from the X type of silicon to the L type of
germanium.}
\label{fig:germanium}
\end{figure}

We show the results for differing Ge concentrations in the active region of
the well in Fig. \ref{fig:germanium}. In the Ge-poor regime $x<0.85$, the
relaxation rate $1/T_{1}$ decreases fairly slowly. Thus designs with a
Si-rich well will not suffer from very fast spin relaxation on the minority
Ge sites. \ 

By contrast, if we cross the critical point around $x=0.85$ where the
conduction band minima switch positions, then there is a very rapid decrease
of $T_{1}$. \ This is due to the much greater g-factor anisotropy of the
[111] minima: $g^{\prime }\approx -0.4$, (as compared to $g^{\prime }<0.001$
in Si).\ The results for the Ge-rich region are much simpler than in the
Si-rich region. \ Uniaxial strain does not affect the relative valley
energies or the matrix elements parametrized by \textbf{D}. \ The only
effect of going to the dot from the impurity is to reduce corrections to the
effective mass approximation, which will strongly decrease the relaxation
time. \ Indeed, if we take the impurity result for Ge:P from Ref. \cite%
{hasegawa}, which is $T_{1}=2.3\times 10^{-3}\,s$ and estimate the decrease
in $\Delta E$ as about a factor of 100, then we obtain $T_{1}\sim 10^{-6}$
to $10^{-7}s,$ which is vastly shorter than that of Si-rich structures. \
The physics is the same as that which governs the divergence of the rate in
Fig. \ref{fig:deltac} for the unstrained case,namely the much smaller energy
denominators.

To make contact with Sec. \ref{sec:well} note that if there is Ge in the
barrier regions, as is the case in most SiGe designs, then we will have an
envelope function that weights Si and Ge differently, then $x$ in the VCA
formulas must be replaced by $f^{2}=\sum_{\vec{R}}|f_{\vec{R}}|^{2}$, where $%
|f_{\vec{R}}|^{2}$ is the amplitude that the electron is on a Ge site and
the sum runs only over Ge sites, as described in Sec. \ref{sec:well}. \ This
is the product of the Ge concentration in the layer, times the probability
that the electron is on the layer, summed over layers. \ Referring to Figs. %
\ref{fig:donor} and \ref{fig:qw}, we see that $f^{2}$ is always less than $%
10^{-3}$ for structures that have pure Si active layers and Ge in the
barrier layers. \ The effect on $T_{1}$ is small so these structures are
''phonon-safe''.

\begin{table}[tbp]
\caption{Materials parameters used in this work.}
\label{tab:params}%\begin{ruledtabular}
\begin{tabular}{ccc}
\hline\hline
Parameter & Value & Ref.~[\# ] \\ \hline
$a_{\mathrm{Si}}$ & 5.43 \AA & \cite{schaeffler97} \\ \hline
$a_{\mathrm{Ge}}$ & 5.66 \AA & \cite{schaeffler97} \\ \hline
$c_{11}$ & 1.675 & \cite{schaeffler97} \\ \hline
$c_{12}$ & 0.650 & \cite{schaeffler97} \\ \hline
$\Xi_u^{(\Delta )}$ & 9.29 eV & \cite{rieger93} \\ \hline
$\Xi_d^{(\Delta )}$ & -10.7 eV & \cite{fischetti96} \\ \hline\hline
\end{tabular}
%\end{ruledtabular}
\end{table}

\section{Conclusion}

\label{sec:conclusion}

Decoherence due to spin relaxation by emission and absorption of single
phonons does not pose a major obstacle to quantum computation in quantum dot
SiGe heterostructures using spin qubits, if these qubits are properly
designed. This spin relaxation mechanism is the dominant one in the case of
electrons in donor bound states, such as in Si:P. \ 

In order that this source of decoherence be kept under control, however,
certain conditions must be satisfied in the design of the structure.
Unstrained SiGe alloys generally have relatively short spin relaxation times
due to valley degeneracy. \ This degeneracy produces strong spin mixing in
the Kramers-degenerate ground states of the quantum dot once the field is
turned on. This in turn increases the electron-phonon matrix element that
connects the two states and decreases the relaxation time. \ This effect,
fundamentally due to spin-orbit coupling, is stronger in the quantum dot
than in the impurity state. \ The corrections to effective-mass theory,
stronger for the impurity potential, tend to lift the degeneracy. \ 

The crucial role that strain plays, in Si-rich structures, is to lift the
valley degeneracy. \ In this respect compressive uniaxial strain is the
best, since the residual degeneracy is reduced to two valleys. \ The
wavefunction for the ground state is then symmetric in the valley index, and
the first excited state, the only one with a small energy denominator, is
antisymmetric in this index. \ This fact strongly suppresses the spin mixing
effect of the external field. \ Thus a workable design should include
uniaxial strain that exceeds a critical value. \ This value is small enough
that there will be sufficient strain in most practical structures in which
the active layer is Si rich and sandwiched between relatively Ge-rich
layers. \ The active layer must be thin enough to avoid dislocation-mediated
relaxation.

Spin-orbit coupling is stronger in Ge than in Si due to the higher atomic
number. \ This suggests that Si-rich structures are to be preferred, and the
calculations bear this out. \ The relaxation time decreases with increasing $%
x$ in a Si$_{1-x}$Ge$_{x}$ alloy. \ In fact, if $x$ increases beyond $x=0.85$%
, decoherence becomes very strong, and quantum dots in these Ge-rich
structures, such as those proposed by Vrijen\textit{\ et al. \cite{vrijen}, }%
may well run into difficulties for this reason. \ A key point for these
structures is that uniaxial stress does not lift the valley degeneracy,
since the valleys have moved off the crystal axes. \ Strain-induced
suppression of spin relaxation cannot occur.

Beyond design issues, fabrication quality is also important. \ Even a small
concentration of magnetic impurities will negatively impact the spin
relaxation. \ In modern semiconductor technology, however, concentrations of
magnetic impurities much less than 0.1 ppm are routine. \ The impact of
lattice imperfections is less clear. \ These will generally act to lower the
symmetry of the system, and to lessen the accuracy of the effective-mass
approximation. \ As we have seen in the context of the impurity
calculations, these effects generally increase the spin relaxation time
because of reduced degeneracy. \ On the other hand, localized bound states
can form at such imperfections. \ If this results in free electron spins, it
could have seriously negative effects on the relaxation time. Similar
effects would result from lattice defects that produce localized phonon
modes.

Finally, it appears that quantum dot designs will need rather low
temperatures in order to operate. \ The impurity experiments clearly
indicate that multi-phonon relaxation increase rapidly with temperature \cite%
{tgcastner}, with a crossover to this regime at about 3 K. \ We speculate
that this crossover temperature is pushed up when the effects of strain are
included, but we have not yet done the requisite calculations.

We would like to acknowledge useful discussions with our collaborators at
the University of Wisconsin-Madison and with Xuedong Hu. \ This work was
supported by the National Science Foundation under the Materials Theory
Program, Grant No. DMR-0081039, the Materials Research Science and
Engineering Center Program, Grant No. DMR-0079983, and the Army Research
Office.

\end{document}